\RequirePackage[2020-02-02]{latexrelease}
\documentclass[preprint,double-spaced,showpacs,preprintnumbers,amsmath,amssymb]{revtex4}

\usepackage{graphicx}
\usepackage{dcolumn}
\usepackage{bm,slashed}

\begin{document}

\preprint{}

\title{The Dirac equation as one fourth-order equation for one function -- a general, manifestly covariant form}

\author{Andrey Akhmeteli}
 \email{akhmeteli@ltasolid.com}
 \homepage{http://www.akhmeteli.org}
\affiliation{%
LTASolid Inc.\\
10616 Meadowglen Ln 2708\\
Houston, TX 77042, USA}


\date{\today}

\begin{abstract}
 Previously (A. Akhmeteli, J. Math. Phys., v. 52, p. 082303 (2011)), the Dirac equation in an arbitrary electromagnetic field was shown to be generally equivalent to a fourth-order equation for just one component of the four-component Dirac spinor function. This was done for a specific (chiral) representation of gamma-matrices and for a specific component.  In the current work, the result is generalized for a general representation of gamma-matrices and a general component (satisfying some conditions). The resulting equivalent of the Dirac equation is also manifestly relativistically covariant and should be useful in applications of the Dirac equation.
\end{abstract}

\maketitle

\section{\label{sec:level1a}Introduction}

The Dirac equation ``remains a cornerstone of physics to this day''~\cite{Wil}. It is crucial for such diverse areas as high energy physics and quantum chemistry (e.g., it is required to explain the properties of the ubiquitous lead-acid batteries~\cite{Ahuja}).

Dirac sought an equation of the first order in time~\cite{Dirac28}. To this end, he had to introduce a four-component spinor function. Feynman and Gell-Mann~\cite{Feygel} argued that the wave function does not have to have four complex components and showed that the Dirac equation is equivalent to a second-order equation for a two-component function. It was shown recently (~\cite{Akhmeteli-JMP}; see also ~\cite{Bagrov2014}, pp. 24-25) that, surprisingly, in a general case, three out of four complex components of the Dirac spinor can be algebraically eliminated from the Dirac equation in an arbitrary electromagnetic field. Therefore, the Dirac equation is generally equivalent to a fourth-order partial differential equation for just one component, which can be made real (at least locally) by a gauge transform. However, this result was derived for a specific (chiral) representation of $\gamma$-matrices and for a specific component. In this article, the fourth-order equation for one function, which is equivalent to the Dirac equation, is derived for an arbitrary set of $\gamma$-matrices satisfying the standard hermiticity conditions and for an arbitrary component that is also a component of the right-handed or the left-handed part of the Dirac spinor function. The resulting equation is also manifestly relativistically covariant, unlike that of  Ref.~\cite{Akhmeteli-JMP}. This non-trivial result adds to the immense beauty of the Dirac equation and belongs in textbooks. It is important both for foundations of quantum theory (see ~\cite{Akhmeteli-IJQI},~\cite{Akhmeteli-EPJC}) and for numerous applications of the Dirac equation.

\maketitle

\section{\label{sec:level1b}Algebraic elimination of components from the Dirac equation in a general form}

Let us start with the Dirac equation in the following form:
\begin{equation}\label{eq:pr25}
(i\slashed{\partial}-\slashed{A})\psi=\psi,
\end{equation}
where, e.g., $\slashed{A}=A_\mu\gamma^\mu$ (the Feynman slash notation). For the sake of simplicity, a system of units $\hbar=c=m=1$ is used, and the electric charge $e$ is included in $A_\mu$ ($eA_\mu \rightarrow A_\mu$). The metric tensor used to raise and lower indices is ~\cite{Itzykson}
\begin{equation}\label{eq:d1ps}
\nonumber
g_{\mu\nu}=g^{\mu\nu}=\left( \begin{array}{cccc}
1 & 0 & 0 & 0\\
0 & -1 & 0 & 0\\
0 & 0 & -1 & 0\\
0 & 0 & 0 & -1 \end{array} \right), g_\mu^\nu=\delta_\mu^\nu.
\end{equation}
Multiplying both sides of equation (\ref{eq:pr25}) by $(i\slashed{\partial}-\slashed{A})$ from the left and using notation
\begin{equation}\label{eq:li2}
\sigma^{\mu\nu}=\frac{i}{2}[\gamma^\mu,\gamma^\nu],
\end{equation}
\begin{eqnarray}\label{eq:d8n1}
F^{\mu\nu}=A^{\nu,\mu}-A^{\mu,\nu}=\left( \begin{array}{cccc}
0 & -E^1 & -E^2 & -E^3\\
E^1 & 0 & -H^3 & H^2\\
E^2 & H^3 & 0 & -H^1\\
E^3 &-H^2 & H^1 & 0  \end{array} \right),
\end{eqnarray}
we obtain:
\begin{eqnarray}\label{eq:li1}
\nonumber
\psi=(i\gamma^\nu\partial_\nu-A_\nu\gamma^\nu)(i\gamma^\mu\partial_\mu-A_\mu\gamma^\mu)\psi=
\\
\nonumber
(-\gamma^\nu\gamma^\mu\partial_\nu\partial_\mu-i A_\nu\gamma^\nu\gamma^\mu\partial_\mu-i\gamma^\nu A_\mu\gamma^\mu\partial_\nu-i\gamma^\nu\gamma^\mu A_{\mu,\nu}+A_\nu A_\mu\gamma^\nu\gamma^\mu)\psi=
\\
\nonumber
(-\frac{1}{2}(\gamma^\nu\gamma^\mu+\gamma^\mu\gamma^\nu)\partial_\nu\partial_\mu-i A_\nu\gamma^\nu\gamma^\mu\partial_\mu-i\gamma^\mu A_\nu\gamma^\nu\partial_\mu-i\gamma^\nu\gamma^\mu A_{\mu,\nu}+
\\
\nonumber
\frac{1}{2}(A_\nu A_\mu\gamma^\nu\gamma^\mu+A_\mu A_\nu\gamma^\mu\gamma^\nu))\psi=
\\
\nonumber
(-g^{\mu\nu}\partial_\nu\partial_\mu-2 i A_\nu g^{\mu\nu}\partial_\mu-\frac{i}{2}(\gamma^\nu\gamma^\mu A_{\mu,\nu}+\gamma^\mu\gamma^\nu A_{\nu,\mu})+A_\nu A_\mu g^{\mu\nu})\psi=
\\
\nonumber
(-\partial^\mu\partial_\mu-2 i A^\mu\partial_\mu-\frac{i}{2}(\gamma^\nu\gamma^\mu A_{\mu,\nu}+(2 g^{\mu\nu}-\gamma^\nu\gamma^\mu) A_{\nu,\mu})+A^\mu A_\mu)\psi=
\\
\nonumber
(-\partial^\mu\partial_\mu-2 i A^\mu\partial_\mu-\frac{i}{2}(\gamma^\nu\gamma^\mu (A_{\mu,\nu}-A_{\nu,\mu})+2 A^\mu_{,\mu})+A^\mu A_\mu)\psi=
\\
\nonumber
(-\partial^\mu\partial_\mu-2 i A^\mu\partial_\mu-\frac{i}{4}(\gamma^\nu\gamma^\mu-\gamma^\mu\gamma^\nu)F_{\nu\mu}-i A^\mu_{,\mu}+A^\mu A_\mu)\psi=
\\
(-\partial^\mu\partial_\mu-2 i A^\mu\partial_\mu-i A^\mu_{,\mu}+A^\mu A_\mu-\frac{1}{2}F_{\nu\mu}\sigma^{\nu\mu})\psi.
\end{eqnarray}
(A similar equation can be found in the original article by Dirac ~\cite{Dirac28}. Feynman and Gell-Mann~\cite{Feygel} used a similar equation to eliminate two out of four components of the Dirac spinor function). We obtain:
\begin{equation}\label{eq:li3}
(\Box'+F)\psi=0,
\end{equation}
where the modified d'Alembertian $\Box'$ is defined as follows:
\begin{eqnarray}\label{eq:d8n2}
\Box'=\partial^\mu\partial_\mu+2 i A^\mu\partial_\mu+i A^\mu_{,\mu}-A^\mu A_\mu+1=-(i\partial_\mu-A_\mu)(i\partial^\mu-A^\mu)+1,
\end{eqnarray}
and
\begin{equation}\label{eq:li4}
F=\frac{1}{2}F_{\nu\mu}\sigma^{\nu\mu}.
\end{equation}
Let us note that $\Box'$ and $F$ are manifestly relativistically covariant.

We assume that the set of $\gamma$-matrices satisfies the standard hermiticity conditions ~\cite{Itzykson}:
\begin{equation}\label{eq:li5a}
\gamma^{\mu\dag}=\gamma^0\gamma^\mu\gamma^0, \gamma^{5\dag}=\gamma^5.
\end{equation}
Then a charge conjugation  matrix $C$ can be chosen in such a way (~\cite{Bogol},~\cite{Schweber}) that
\begin{equation}\label{eq:li5}
C\gamma^\mu C^{-1}=-\gamma^{\mu T}, C\gamma^5 C^{-1}=\gamma^{5 T}, C\sigma^{\mu\nu}C^{-1}=-\sigma^{\mu\nu T},
\end{equation}
\begin{equation}\label{eq:li6}
C^T=C^\dag=-C, CC^\dag=C^\dag C=I, C^2=-I,
\end{equation}
where the superscript $T$ denotes transposition, and $I$ is the unit matrix.

Let us choose a component of the Dirac spinor $\psi$ in the form $\bar{\xi}\psi$, where $\xi$ is a constant spinor  (so it does not depend on the spacetime coordinates $x=(x^0,x^1,x^2,x^3)$, and $\partial_\mu\xi\equiv 0$), and multiply  both sides of equation (\ref{eq:li3}) by $\bar{\xi}$ from the left:
\begin{equation}\label{eq:li7}
\Box'(\bar{\xi}\psi)+\bar{\xi}F\psi=0.
\end{equation}
To derive an equation for only one component $\bar{\xi}\psi$, we need to express $\bar{\xi}F\psi$ via $\bar{\xi}\psi$, but the author cannot do this for an arbitrary spinor $\xi$ (or prove that this cannot be done). Therefore, to simplify this task, we demand that $\xi$ is an eigenvector of $\gamma^5$ (in other words, $\xi$ is either right-handed or left-handed). This condition is Lorentz-invariant. Indeed, Dirac spinors $\chi$ transform under a Lorentz transformation as follows:
\begin{equation}\label{eq:li7a}
\chi'=\Lambda \chi,
\end{equation}
where matrix $\Lambda$ is non-singular and commutes with $\gamma^5$ if the Lorentz transformation is proper and anticommutes otherwise ~\cite{Bogo}. Therefore, if $\xi$ is an eigenvector of $\gamma^5$, then $\xi'$ is also an eigenvector of $\gamma^5$, although not necessarily with the same eigenvalue.

Eigenvalues of $\gamma^5$ equal either $+1$ or $-1$, so $\gamma^5\xi=\pm\xi$. The linear subspace of eigenvectors of $\gamma^5$ with the same eigenvalue as $\xi$ is two-dimensional, so we can choose another constant spinor $\eta$ that is an eigenvector of $\gamma^5$  with the same eigenvalue as $\xi$ in such a way that $\xi$ and $\eta$ are linearly independent. This choice is Lorentz-covariant, as matrix $\Lambda$ in equation (\ref{eq:li7a}) is non-singular.

Obviously, we can derive an equation similar to (\ref{eq:li7}) for $\eta$:
\begin{equation}\label{eq:li7f}
\Box'(\bar{\eta}\psi)+\bar{\eta}F\psi=0.
\end{equation}

If $\gamma^5\xi=\pm\xi$, then $\bar{\xi}=\xi^\dag\gamma^0$ is a left eigenvector of $\gamma^5$ with an eigenvalue $\mp 1$, as
\begin{equation}\label{eq:li7b}
\bar{\xi}\gamma^5=\xi^\dag\gamma^0\gamma^5=-\xi^\dag\gamma^5\gamma^0=-(\gamma^5\xi)^\dag\gamma^0=\mp\bar{\xi}.
\end{equation}
The same is true for spinors $\bar{\eta}=\eta^\dag\gamma^0$ (the proof is identical to that in (\ref{eq:li7b})), $\bar{\xi}F$, and $\bar{\eta}F$, as $\gamma^5$ commutes with $\sigma^{\mu\nu}$ ~\cite{Itzykson}. As the subspace of left eigenvectors of $\gamma^5$ with an eigenvalue $\mp 1$ is two-dimensional and includes spinors $\bar{\xi}F$, $\bar{\eta}F$, $\bar{\xi}$, and $\bar{\eta}$, where the two latter spinors are linearly independent (otherwise spinors $\xi$ and $\eta$ would not be linearly independent), there exist such $a=a(x)$, $b=b(x)$, $a'=a'(x)$, $b'=b'(x)$ that
\begin{equation}\label{eq:lin12}
\bar{\xi}F=a\bar{\xi}+b\bar{\eta},
\end{equation}
\begin{equation}\label{eq:lin13}
\bar{\eta}F=a'\bar{\xi}+b'\bar{\eta}.
\end{equation}

For each spinor $\chi$ the charge conjugated spinor
\begin{equation}\label{eq:lin14}
\chi^c=C\bar{\chi}^T
\end{equation}
can be defined, and it has the same transformation properties under Lorentz transformations as $\chi$ ~\cite{Schweber}. We have
\begin{equation}\label{eq:lin15}
\bar{\chi}\chi^c=\bar{\chi}C\bar{\chi}^T=(\bar{\chi})_\alpha C_{\alpha\beta}(\bar{\chi})_\beta=0,
\end{equation}
as $(\bar{\chi})_\alpha(\bar{\chi})_\beta$ and $ C_{\alpha\beta}$ are respectively symmetric and antisymmetric (see equation (\ref{eq:li6})) with respect to transposition of $\alpha$ and $\beta$.

Let us multiply equations (\ref{eq:lin12}),(\ref{eq:lin13}) by $\xi^c$ and $\eta^c$ from the right:
\begin{eqnarray}\label{eq:lin16}
\nonumber
\bar{\xi}F\xi^c=a(\bar{\xi}\xi^c)+b(\bar{\eta}\xi^c)=b(\bar{\eta}\xi^c),
\\
\nonumber
\bar{\xi}F\eta^c=a(\bar{\xi}\eta^c)+b(\bar{\eta}\eta^c)=a(\bar{\xi}\eta^c),
\\
\nonumber
\bar{\eta}F\xi^c=a'(\bar{\xi}\xi^c)+b'(\bar{\eta}\xi^c)=b'(\bar{\eta}\xi^c),
\\
\nonumber
\bar{\eta}F\eta^c=a'(\bar{\xi}\eta^c)+b'(\bar{\eta}\eta^c)=a'(\bar{\xi}\eta^c),
\end{eqnarray}
so
\begin{eqnarray}\label{eq:lin17}
a=\frac{\bar{\xi}F\eta^c}{\bar{\xi}\eta^c},
b=\frac{\bar{\xi}F\xi^c}{\bar{\eta}\xi^c},
a'=\frac{\bar{\eta}F\eta^c}{\bar{\xi}\eta^c},
b'=\frac{\bar{\eta}F\xi^c}{\bar{\eta}\xi^c}.
\end{eqnarray}
Let us note that
\begin{equation}\label{eq:lin18}
\bar{\xi}\eta^c=\bar{\xi}C\bar{\eta}^T=(\bar{\xi})_\alpha C_{\alpha\beta}(\bar{\eta})_\beta=-(\bar{\eta})_\beta C_{\beta\alpha}(\bar{\xi})_\alpha=-\bar{\eta}\xi^c
\end{equation}
and
\begin{equation}\label{eq:lin19}
\bar{\xi}F\eta^c=\bar{\xi}F C\bar{\eta}^T=(\bar{\xi}F C\bar{\eta}^T)^T=\bar{\eta}C^T F^T\bar{\xi}^T=\bar{\eta}F C\bar{\xi}^T=\bar{\eta}F\xi^c,
\end{equation}
as
\begin{equation}\label{eq:lin20}
\sigma_{\mu\nu}C=-C\sigma_{\mu\nu}^T
\end{equation}
(see equations (\ref{eq:li5}),(\ref{eq:li6})). Therefore,
\begin{equation}\label{eq:lin20a}
b'=-a.
\end{equation}

Equations (\ref{eq:li7}), (\ref{eq:li7f}),(\ref{eq:lin12}), (\ref{eq:lin13}) yield
\begin{eqnarray}\label{eq:lin21}
\nonumber
\Box'(\bar{\xi}\psi)+a(\bar{\xi}\psi)+b(\bar{\eta}\psi)=0,
\\
\nonumber
\Box'(\bar{\eta}\psi)+a'(\bar{\xi}\psi)+b'(\bar{\eta}\psi)=\Box'(\bar{\eta}\psi)+a'(\bar{\xi}\psi)-a(\bar{\eta}\psi)=0,
\end{eqnarray}
so
\begin{equation}\label{eq:lin22}
\bar{\eta}\psi=-b^{-1}(\Box'(\bar{\xi}\psi)+a(\bar{\xi}\psi))
\end{equation}
and
\begin{equation}\label{eq:lin23}
(\Box'-a)(-b^{-1})(\Box'+a)(\bar{\xi}\psi)+a'(\bar{\xi}\psi)=0
\end{equation}
or
\begin{equation}\label{eq:lin24}
((\Box'-a)b^{-1}(\Box'+a)-a')(\bar{\xi}\psi)=0.
\end{equation}
Substituting the expressions for $a$, $b$, $a'$ from equation (\ref{eq:lin17}) and using equation (\ref{eq:lin18}), we finally obtain:
\begin{equation}\label{eq:lin25}
(((\bar{\xi}\eta^c)\Box'-\bar{\xi}F\eta^c)(\bar{\xi}F\xi^c)^{-1}((\bar{\xi}\eta^c)\Box'
+\bar{\xi}F\eta^c)+\bar{\eta}F\eta^c)(\bar{\xi}\psi)=0.
\end{equation}
This equation looks more complex than equation (21) of Ref.~\cite{Akhmeteli-JMP}, but it is much more general and manifestly relativistically covariant. It can be slightly simplified if we require that $\xi$ and $\eta$ are normalized in such a way that $\bar{\xi}\eta^c=1$ (according to equation (\ref{eq:lin15}), this condition also implies linear independence of $\xi$ and $\eta$). Further simplification can be achieved using the following notation for components of the electromagnetic field:
\begin{equation}\label{eq:simp}
f_{\alpha \beta}=\bar{\alpha}F\beta^c,
\end{equation}
where $\alpha$ and $\beta$ are some Dirac spinors. Then we obtain the following instead of equation (\ref{eq:lin25}):
\begin{equation}\label{eq:sim2}
((\Box'-f_{\xi\eta})f_{\xi\xi}^{-1}(\Box'
+f_{\xi\eta})+f_{\eta\eta})(\bar{\xi}\psi)=0.
\end{equation}

\section{\label{sec:level1b2}Equivalency to the Dirac equation}

Let us first prove that a different choice of $\eta$ yields an equivalent equation. As the subspace of eigenvectors of $\gamma^5$ with the same eigenvalue as $\xi$ is two-dimensional, the different choice $\eta'$ can be expressed as follows:
\begin{equation}\label{eq:lin34}
\eta'=\sigma \eta+\tau\xi,
\end{equation}
where $\sigma$ and $\tau$ are constant, and $\sigma\neq 0$, as otherwise $\eta'$ and $\xi$ would not be linearly independent. We need to substitute $\eta$ in the operator acting on $\bar{\xi}\psi$ in equation (\ref{eq:lin25}) with the expression for $\eta'$ from equation (\ref{eq:lin34}), but let us first note that
\begin{eqnarray}\label{eq:lin35}
\bar{\xi}\eta'^c=\bar{\xi}(\sigma^* \eta^c+\tau^*\xi^c)=\sigma^*(\bar{\xi}\eta^c),
\\
\bar{\xi}F\eta'^c=\sigma^* (\bar{\xi}F\eta^c)+\tau^*(\bar{\xi}F\xi^c),
\\
\bar{\eta'}F\eta'^c=(\sigma^*)^2(\bar{\eta}F\eta^c)+2\sigma^*\tau^*(\bar{\xi}F\eta^c)+(\tau^*)^2(\bar{\xi}F\xi^c)
\end{eqnarray}
(we used equations (\ref{eq:lin15},\ref{eq:lin19})). The substitution then yields
\begin{eqnarray}\label{eq:lin36}
((\bar{\xi}\eta'^c)\Box'-\bar{\xi}F\eta'^c)(\bar{\xi}F\xi^c)^{-1}((\bar{\xi}\eta'^c)\Box'
+\bar{\xi}F\eta'^c)+\bar{\eta'}F\eta'^c=
\\
\nonumber
(\sigma^*(\bar{\xi}\eta^c)\Box'-\sigma^*(\bar{\xi}F\eta^c)-\tau^*(\bar{\xi}F\xi^c))(\bar{\xi}F\xi^c)^{-1}
(\sigma^*(\bar{\xi}\eta^c)\Box'+\sigma^*(\bar{\xi}F\eta^c)+\tau^*(\bar{\xi}F\xi^c))+
\\
\nonumber
(\sigma^*)^2(\bar{\eta}F\eta^c)+2\sigma^*\tau^*(\bar{\xi}F\eta^c)+(\tau^*)^2(\bar{\xi}F\xi^c)=
\\
\nonumber
(\sigma^*)^2
((\bar{\xi}\eta^c)\Box'-\bar{\xi}F\eta^c)(\bar{\xi}F\xi^c)^{-1}((\bar{\xi}\eta^c)\Box'+\bar{\xi}F\eta^c)-
\sigma^*\tau^*((\bar{\xi}\eta^c)\Box'+\bar{\xi}F\eta^c)+
\\
\nonumber
\sigma^*\tau^*((\bar{\xi}\eta^c)\Box'-\bar{\xi}F\eta^c)-(\tau^*)^2(\bar{\xi}F\xi^c)+(\sigma^*)^2(\bar{\eta}F\eta^c)+
2\sigma^*\tau^*(\bar{\xi}F\eta^c)+(\tau^*)^2(\bar{\xi}F\xi^c)=
\\
\nonumber
(\sigma^*)^2
(((\bar{\xi}\eta^c)\Box'-\bar{\xi}F\eta^c)(\bar{\xi}F\xi^c)^{-1}((\bar{\xi}\eta^c)\Box'+\bar{\xi}F\eta^c)+
\bar{\eta}F\eta^c).
\end{eqnarray}
Thus, the operator after the substitution coincides with the original one up to a constant factor, so equation (\ref{eq:lin25}) does not  depend on the choice of $\eta$.

This equation for one component $\bar{\xi}\psi$ is generally equivalent to the Dirac equation (if $\bar{\xi}F\xi^c\slashed{\equiv}0$): on the one hand, it was derived from the Dirac equation, on the other hand, the Dirac spinor $\psi$ can be restored if its component $\bar{\xi}\psi$ is known (a more precise definition of the equivalency is provided below, after equation (\ref{eq:lin46})). Let us demonstrate that.

If $\bar{\xi}\psi$ is known, another component, $\bar{\eta}\psi$, can be determined using equation (\ref{eq:lin22}). Then $\psi$ can be expressed as a sum of a right-handed and a left-handed spinors $\psi_+$ and $\psi_-$, where $\gamma^5 \psi_\pm=\pm\psi_\pm$:
\begin{eqnarray}\label{eq:lin26}
\psi=\psi_++\psi_-,
\\
\psi_\pm=\frac{1}{2}(1\pm\gamma^5)\psi.
\end{eqnarray}
Then $\psi_\mp$ can be expressed as a linear combination of $\xi^c$ and $\eta^c$ (one can show that these two spinors are also eigenvectors of $\gamma^5$ with an eigenvalue $\mp 1$ and are linearly independent, and the subspace of eigenvectors of $\gamma^5$ with an eigenvalue $\mp 1$ is 2-dimensional):
\begin{equation}\label{eq:lin27}
\psi_\mp=u\xi^c+v\eta^c,
\end{equation}
where $u=u(x)$ and $v=v(x)$.

Let us note that, e.g.,
\begin{equation}\label{eq:lin28}
\bar{\xi}\psi=\bar{\xi}\psi_\pm+\bar{\xi}\psi_\mp=\bar{\xi}\psi_\mp,
\end{equation}
as
\begin{equation}\label{eq:lin29}
\bar{\xi}\psi_\pm=\frac{1}{2}\bar{\xi}(1\pm\gamma^5)\psi=0
\end{equation}
($\bar{\xi}$ is a left eigenvector of $\gamma^5$ with an eigenvalue $\mp 1$).
Therefore, we  can multiply equation (\ref{eq:lin27}) by $\bar{\xi}$ and $\bar{\eta}$ from the left:
\begin{eqnarray}\label{eq:lin30}
\nonumber
\bar{\xi}\psi=\bar{\xi}\psi_\mp=u(\bar{\xi}\xi^c)+v(\bar{\xi}\eta^c)=v(\bar{\xi}\eta^c),
\\
\bar{\eta}\psi=\bar{\eta}\psi_\mp=u(\bar{\eta}\xi^c)+v(\bar{\eta}\eta^c)=u(\bar{\eta}\xi^c)
\end{eqnarray}
(we took into account equation (\ref{eq:lin15})).

Thus, $\psi_\mp$ can be expressed via components $\bar{\xi}\psi$ and $\bar{\eta}\psi$ as follows:
\begin{equation}\label{eq:lin31}
\psi_\mp=\frac{(\bar{\xi}\psi)\eta^c-(\bar{\eta}\psi)\xi^c}{\bar{\xi}\eta^c}
\end{equation}
(note equation (\ref{eq:lin18})).
When $\psi_\mp$ is found in this way, $\psi_{\pm}$ can be found using the Dirac equation (\ref{eq:pr25}):
\begin{equation}\label{eq:lin32}
(i\slashed{\partial}-\slashed{A})\psi_\mp=\frac{1}{2}(i\slashed{\partial}-\slashed{A})(1\mp\gamma^5)\psi=
\frac{1}{2}(1\pm\gamma^5)(i\slashed{\partial}-\slashed{A})\psi=\psi_\pm,
\end{equation}
thus, the Dirac spinor can be fully restored if component $\bar{\xi}\psi$ is known.

Let us explicitly prove that the expression for $\psi_\mp$ (equation (\ref{eq:lin31})) and, therefore, the expression for $\psi$ (equation (\ref{eq:lin26})) do not depend on the choice of $\eta$. We have from equations (\ref{eq:lin17},\ref{eq:lin18},\ref{eq:lin22}):
\begin{equation}\label{eq:lin37}
\bar{\eta}\psi=(\bar{\xi}F\xi^c)^{-1}((\bar{\xi}\eta^c)\Box'+\bar{\xi}F\eta^c)(\bar{\xi}\psi),
\end{equation}
therefore, we obtain from equation (\ref{eq:lin31}):
\begin{equation}\label{eq:lin38}
\psi_\mp=(\bar{\xi}\eta^c)^{-1}((\bar{\xi}\psi)\eta^c-
(\bar{\xi}F\xi^c)^{-1}((\bar{\xi}\eta^c)\Box'+\bar{\xi}F\eta^c)(\bar{\xi}\psi)\xi^c).
\end{equation}
Substituting $\eta$ in equation (\ref{eq:lin38}) with $\eta'$ (see equation (\ref{eq:lin34})), we obtain, using equations (29,30):
\begin{eqnarray}\label{eq:lin39}
\nonumber
\psi_\mp=(\sigma^*\bar{\xi}\eta^c)^{-1}\times
\\
\nonumber
((\bar{\xi}\psi)(\sigma^*\eta^c+\tau^*\xi^c)-
(\bar{\xi}F\xi^c)^{-1}(\sigma^*(\bar{\xi}\eta^c)\Box'+\sigma^*(\bar{\xi}F\eta^c)
+\tau^*(\bar{\xi}F\xi^c))(\bar{\xi}\psi)\xi^c)=
\\
(\bar{\xi}\eta^c)^{-1}((\bar{\xi}\psi)\eta^c-
(\bar{\xi}F\xi^c)^{-1}((\bar{\xi}\eta^c)\Box'+\bar{\xi}F\eta^c)(\bar{\xi}\psi)\xi^c).
\end{eqnarray}
Therefore, the expression for $\psi$ defined by equations (\ref{eq:lin26},\ref{eq:lin31},\ref{eq:lin32}) does not depend on the choice of $\eta$.

Let us prove that $\psi$ defined by equations (\ref{eq:lin26},\ref{eq:lin31},\ref{eq:lin32}) satisfies the Dirac equation (\ref{eq:pr25}). This is not quite obvious as the set of solutions of equation (\ref{eq:li3}) used to derive equation (\ref{eq:lin25}) is broader than the set of solutions of the Dirac equation (additional solutions appeared as a result of multiplication by $(i\slashed{\partial}-\slashed{A})$; as a result, equation (\ref{eq:li3}) does not require the right-handed and left-handed parts of a solution to be related -- cf. ~\cite{Feygel}). To prove that $\psi$ satisfies the Dirac equation, it is sufficient to prove that
\begin{equation}\label{eq:lin40}
(i\slashed{\partial}-\slashed{A})\psi_\pm=\psi_\mp,
\end{equation}
as that would imply that
\begin{equation}\label{eq:lin41}
(i\slashed{\partial}-\slashed{A})\psi=(i\slashed{\partial}-\slashed{A})(\psi_\pm+\psi_\mp)=\psi_\mp+\psi_\pm=\psi
\end{equation}
(note equation (\ref{eq:lin32})). Equation (\ref{eq:lin40}) is equivalent to the following equation:
\begin{equation}\label{eq:lin42}
(i\slashed{\partial}-\slashed{A})(i\slashed{\partial}-\slashed{A})\psi_\mp=\psi_\mp
\end{equation}
(again, note equation (\ref{eq:lin32})), or
\begin{equation}\label{eq:lin43}
(\Box'+F)\psi_\mp=0
\end{equation}
(cf. equations (\ref{eq:li1},\ref{eq:li3})). As $\Box'+F$ commutes with $\gamma^5$, $(\Box'+F)\psi_\mp$ is an eigenvector of $\gamma^5$ with the same eigenvalue $\mp1$ as $\psi_\mp$, thus, it can be presented as a linear combination of $\eta^c$ and $\xi^c$. Therefore, to prove equation (\ref{eq:lin43}), it is sufficient to prove that the coefficients in the linear combination vanish, or, equivalently, that
\begin{equation}\label{eq:lin44}
\bar{\xi}(\Box'+F)\psi_\mp=\bar{\eta}(\Box'+F)\psi_\mp=0
\end{equation}
(cf. equations (\ref{eq:lin30})). Using equations (\ref{eq:lin12},\ref{eq:lin13},\ref{eq:lin15},\ref{eq:lin16},\ref{eq:lin18},\ref{eq:lin21},\ref{eq:lin31}), we obtain:
\begin{eqnarray}\label{eq:lin45}
\nonumber
\bar{\xi}(\Box'+F)\psi_\mp=(\bar{\xi}\Box'+a\bar{\xi}+b\bar{\eta})
((\bar{\xi}\psi)\eta^c-(\bar{\eta}\psi)\xi^c)(\bar{\xi}\eta^c)^{-1}=
\\
(\Box'(\bar{\xi}\psi)(\bar{\xi}\eta^c)+a(\bar{\xi}\psi)(\bar{\xi}\eta^c)-b(\bar{\eta}\psi)(\bar{\eta}\xi^c))
(\bar{\xi}\eta^c)^{-1}=
\Box'(\bar{\xi}\psi)+a(\bar{\xi}\psi)+b(\bar{\eta}\psi)=0
\end{eqnarray}
and
\begin{eqnarray}\label{eq:lin46}
\nonumber
\bar{\eta}(\Box'+F)\psi_\mp=(\bar{\eta}\Box'+a'\bar{\xi}+b'\bar{\eta})
((\bar{\xi}\psi)\eta^c-(\bar{\eta}\psi)\xi^c)(\bar{\xi}\eta^c)^{-1}=
\\
(-\Box'(\bar{\eta}\psi)(\bar{\eta}\xi^c)+a'(\bar{\xi}\psi)(\bar{\xi}\eta^c)
-b'(\bar{\eta}\psi)(\bar{\eta}\xi^c))
(\bar{\xi}\eta^c)^{-1}=
\Box'(\bar{\eta}\psi)+a'(\bar{\xi}\psi)+b'(\bar{\eta}\psi)=0
\end{eqnarray}
(note equations (\ref{eq:lin20a},\ref{eq:lin24})).

We can summarize the above as follows. Equation (\ref{eq:lin25}) for one component $\bar{\xi}\psi$  is equivalent to the Dirac equation (provided that we know $\xi$ and that the component of electromagnetic field $\bar{\xi}F\xi^c$ does not vanish identically) in the following sense: the Dirac equation implies equation (\ref{eq:lin25}), and the latter implies the Dirac equation for the Dirac spinor restored from its component $\bar{\xi}\psi$ using equations (19,\ref{eq:lin22},\ref{eq:lin26},\ref{eq:lin31},\ref{eq:lin32}).

To give a physical interpretation to equation (\ref{eq:lin25}), we need to define the current. The latter equals (up to a constant factor):
\begin{equation}\label{eq:lin33}
j^\mu=\bar{\psi}\gamma^\mu\psi=\overline{\psi_\pm}\gamma^\mu\psi_\pm+\overline{\psi_\mp}\gamma^\mu\psi_\mp,
\end{equation}
as one can show that, e.g., $\overline{\psi_\pm}\gamma^\mu\psi_\mp=0$. Thus, the current can be expressed via component $\bar{\xi}\psi$ using equations (\ref{eq:lin17},\ref{eq:lin22},\ref{eq:lin31},\ref{eq:lin32},\ref{eq:lin33}).

Let us  note that equations (\ref{eq:lin24}) or (\ref{eq:lin25}) reduce to the equation derived in Ref.~\cite{Akhmeteli-JMP} in a specific case. In the chiral representation of $\gamma$-matrices ~\cite{Itzykson}
\begin{equation}\label{eq:d1}
\gamma^0=\left( \begin{array}{cc}
0 & -I\\
-I & 0 \end{array} \right),\gamma^i=\left( \begin{array}{cc}
0 & \sigma^i \\
-\sigma^i & 0 \end{array} \right),\gamma^5=\left( \begin{array}{cc}
I & 0\\
0 & -I \end{array} \right),C=\left( \begin{array}{cc}
-i\sigma^2 & 0\\
0 & i\sigma^2 \end{array} \right),
\end{equation}
where index $i$ runs from 1 to 3, and $\sigma^i$ are the Pauli matrices. One can obtain:
\begin{equation}\label{eq:li25}
F=\left( \begin{array}{cccc}
i F^3 & i F^1+F^2 & 0 & 0\\
i F^1-F^2 & -i F^3 & 0 & 0\\
0 & 0 & -i F^{3*} & -i F^{1*}-F^{2*}\\
0 & 0 & -i F^{1*}+F^{2*} & i F^{3*} \end{array} \right),
\end{equation}
where $F^i=E^i+i H^i$, electric field $E^i$ and magnetic field $H^i$ are defined by equation (\ref{eq:d8n1}).

Let us choose
\begin{equation}\label{eq:li26}
\xi=\left( \begin{array}{c}
0\\
0\\
-1\\
0\end{array}\right),
\eta=\left( \begin{array}{c}
0\\
0\\
0\\
1\end{array}\right),
\end{equation}
then, if $\psi$ has components
\begin{equation}\label{eq:d2}
\psi=\left( \begin{array}{c}
\psi_1\\
\psi_2\\
\psi_3\\
\psi_4\end{array}\right),
\end{equation}
one obtains $\bar{\xi}\psi=\psi_1$, $a=i F^3$, $b=-i F^1-F^2$, $a^{'}=-i F^1+F^2$, and  equation (\ref{eq:lin24}) acquires the same form as in Ref.~\cite{Akhmeteli-JMP}:
\begin{eqnarray}\label{eq:q}
\left(\left(\Box'-i F^3\right)\left(i F^1+F^2\right)^{-1}\left(\Box'+i F^3\right)-i F^1+F^2\right)\psi_1=0.
\end{eqnarray}

\maketitle

\section{\label{sec:level1c}Conclusion}

Building on the results of Ref.~\cite{Akhmeteli-JMP}, we have derived the manifestly covariant fourth-order/one-function equivalent of the Dirac equation for the general case of an arbitrary set of $\gamma$-matrices (satisfying the standard hermiticity conditions) and an arbitrary component of the form $\bar{\xi}\psi$ (where $\psi$ is the four-component spinor function of the Dirac equation and $\xi$ is an arbitrary fixed right eigenvector of $\gamma^5$). This fundamental result should be useful for numerous applications of the Dirac equation.

An anonymous referee summarized a result of Ref.~\cite{Akhmeteli-JMP} as follows: ``the Dirac equation is equivalent to a 4th-order equation for a scalar field''. This author was puzzled by such summary, as the equivalent of the Dirac equation in Ref.~\cite{Akhmeteli-JMP} was written for a component of a spinor, not for a scalar. However, the results of the present work seem to suggest that the Dirac equation is indeed equivalent to a 4th-order equation for a scalar field (e,g., $\bar{\xi}\psi$).

\section*{Acknowledgments}

The author is grateful to V. G. Bagrov, A. V. Gavrilin, A. Yu. Kamenshchik, P. W. Morgan, nightlight, R. Sverdlov, and H. Westman for their interest in this work and valuable remarks.

\end{document}